\documentclass[10pt,a4paper,twoside]{article}
\usepackage[utf8]{inputenc}
\usepackage[polish]{babel}
\usepackage{graphicx}
\usepackage{nicefrac}
\usepackage[left=2cm,right=2cm,top=2cm,bottom=2cm]{geometry}
\author{Lukasz Machura$^{1*}$, Paulina Trybek$^1$, Michal Nowakowski$^2$}
\date{}
\title{Inter-pulse intervals of external anal sphincter surface EMG signals recorded from colorectal cancer patients}
\begin{document}
\maketitle
\noindent $^1$ Division of Computational Physics and Electronics, Institute of Physics, Silesian Center for Education and 
Interdisciplinary Research, Chorzow, Poland
\\
$^2$ Department of Medical Education, Jagiellonian University Medical College, Krakow, Poland
\\
$^*$ lukasz.machura@smcebi.edu.pl

\section*{Abstract}
Intervals between electrical pulses generated by the electrical activity produced by the motor 
units of an external anal sphincter were studied at four time intervals during multimodal rectal 
cancer treatment.
Probability distribution function %of such intervals
does not exhibit significant differences for all considered time intervals. 
It is found that the probability distribution 
rescaled with an average interval time can be described by means of the stretched exponential function
with the threshold dependent scale and shape parameters. 
Interval trains possess rather strong correlations as their shuffled counterparts show exponential 
Poisson like probability distribution. 
Finally the clustering effects were not found as the conditional probability distributions can also 
be described by the exponential function.

\section*{Introduction}
\subsection*{Colorectal cancer}
According to the World Health Organization close to 14 million of new cases and 8 million 
of cancer related deaths were recorded in 2012 \cite{SteWilWHO2014}. 
According to prognoses based on current trends in epidemiology of cancer by 2030 we will reach 
$23.6$ million of new cases per year \cite{bray2012global,ferlay2015cancer}. 
Considering cancer specific survival rates depending on 
health care availability, cancer stage and other influencing factors anything from $12$ to
$89\%$ of diagnosed patients can be cured of cancer. Typically multimodal treatment 
including surgery, radiation and chemotherapy is applied. Each of those modes results with 
specific 
and general side effects, many of which are directly related to anatomically adjacent 
structures. To no surprise defecation disorders described as (Low) Anterior Resection 
Syndrome are frequent \cite{ikeuchi1996clinico} 
and affect anything from $6$ to $80\%$ of patients 
\cite{ikeuchi1996clinico,batignani1991affects,suzuki1980anorectal,lewis1995some}
depending on treatment modalities applied, depth of anastomosis, extent of surgery and many 
others often poorly understood factors.
It is unclear which factors contribute mostly to this very
debilitating condition composed of difficulties with evacuation of stool and/or flatus or 
stool incontinence as well as urgency to defecate or sensory deficits. There are some 
suggestions that innervation injury might play a role \cite{rao1996anterior} but compliance deficits or 
neo-rectum itself or surrounding tissues, sphincter deficiency or radiation induced injury 
have also been possible candidates. Since number of patients are very high and prevalence of 
treatment related sphincter dysfunction is on the rise proper evaluation of innervation of 
external anal sphincter becomes rather crucial.

% For both men and
% women colorectal cancer can be found among 5 most common types of cancer. 
% If detected early 
% and treated adequately it can be cured to a great extent \cite{AmeCanSoc2016}.
% Rectal cancer requires complex multimodal treatment composed of surgery, irradiation 
% and chemotherapy. All those methods of treatment can cause significant 
% stool continence related problems hence proper assessment of anorecatal 
% innervation before and after the treatment can be crucial for prevention 
% and treatment of complications. 

\subsection*{Motor units and surface electromyography}
Currently available methods of assessment of innervation of External Anal Sphincter including 
Pudendal Nerve Terminal Motor Latency (PNTML) \cite{hill2002pudendal}
or needle EMG \cite{floyd1953electromyography} turned out not to be sufficiently effective 
\cite{daube2009needle,remes2008neurophysiological}. 
PNTML measures only conduction velocity while needle EMG is invasive, painful and very prone to 
sampling error. Surface EMG (sEMG) offers a possibility to avoid some of limitations hindering 
other methods but has its own limitations
\cite{de1997use,merletti2004multichannel,cescon2011geometry}. 

The smallest functional part of neuromuscular system is called motor unit (MU). It is composed of motor neuron and muscle fibers connected by neuromuscular junctions. The basis of registration EMG signal is detection of action potential and the cycle of depolarization - repolarization generated at the muscle fiber membrane. The wave of depolarization creates an electric dipole moving along the muscle fiber, which can be recorded by external electromyographic devices. The representation of time of action potential generated by MU, so-called motor unit action potential (MUAP) can be described by several parameters including an amplitude or a peak's area or MUAP duration \cite{rodriguez2012motor}. Although the identification of the individual MUAP can provide valuable information about MU recruitment, the proper interpretation of EMG signal is still a serious challenge. 

sEMG is an alternative technique for intra-muscular (needle) electromyography, due to its noninvasive character. It offers specific advantages but also implies several challenges due to its complex character. sEMG does not require the placement of electrodes in the direct proximity to the registered signal sources. Therefore, the relevant advantage of sEMG is the ability to monitor a significant part of the muscle or the whole group of MUs. On the other hand this is the cause of intensification of negative phenomena, consisting of inter-muscular cross-talk \cite{winter1994crosstalk}, other tissue interposition or distance related changes in signal amplitude. The amplitude of the signal is strongly influenced by several factors. First one is the aforementioned effect of cross talk which not only adds unintended signal but also adds or subtracts from summative sEMG signal. Another one is noise generated by external devices. On top of that any modification of the source position in relation to the electrodes might influence amplitude or shape of the signal curve. 

In order to overcome the problem of variation of the amplitude registered it is reasonable to normalize the data to some physiologically meaningful reference value \cite{konrad2005abc}. The most common technique of normalization, especially in kinesiological sEMG is based on the Maximum Voluntary Contraction (MVC), where the maximum Root Mean Square value (RMS) is used to normalize EMG data series. In order to detect the baseline, the relaxation state can be recorded, which also depends of several factors such as the level of external noise and the recording conditions in general. 

% \subsection*{Distribution of the inter--pulse intervals LUB Weibull function}
% Opisac papiery DeLuca, napisac, ze u nas EMG jest inne, miesnie sa inne, nie da sie kilogramow 
% podczepic. DeLuca ma Weibuila - my fitujemy strechd exponential.
% You may title this section "Methods" or "Models". 
% "Models" is not a valid title for PLoS ONE authors. However, PLoS ONE
% authors may use "Analysis" 
\section*{Materials and Methods}
Reported nonstationarities of the EMG signals \cite{DeLuca1973} has not been
replicated in our experiment. The most probable reasons for the nonstationarity in 
the aforementioned case is (a) the use of the invasive needle EMG
technique and (b) recordings at the states of constant force isometric contractions.
On the contrary the signals presented here and recorded with sEMG at both MVC and at rest 
turn out to be stationary for all stages $D_1$ -- $D_4$ of medical treatment. 
Due to the differential detection of the signal in the presented set--up, all signals
tend to be symmetric and the global average value of the voltage is zero
in all cases. In that case the $0\mu V$ value also stands for the baseline signal level. 
%Differential signals are ...
%
% CONSIDER SUPPORTING INFORMATION TXT ON STATIONARITY
%

\subsection*{Point process}
In order to describe the statistical properties of the intervals between the consecutive
electrical activities of MU a point process is built on top of the acquired data.
Before the construction, the original data was scaled with the standard deviation $\sigma_V$. 
This allow for the thresholds $q$ to be expressed in units of $\sigma_V$. This scaling allows for 
the comparison of signals with volatilities of different relative values.
\begin{figure}[htbp]
\includegraphics[scale=1]{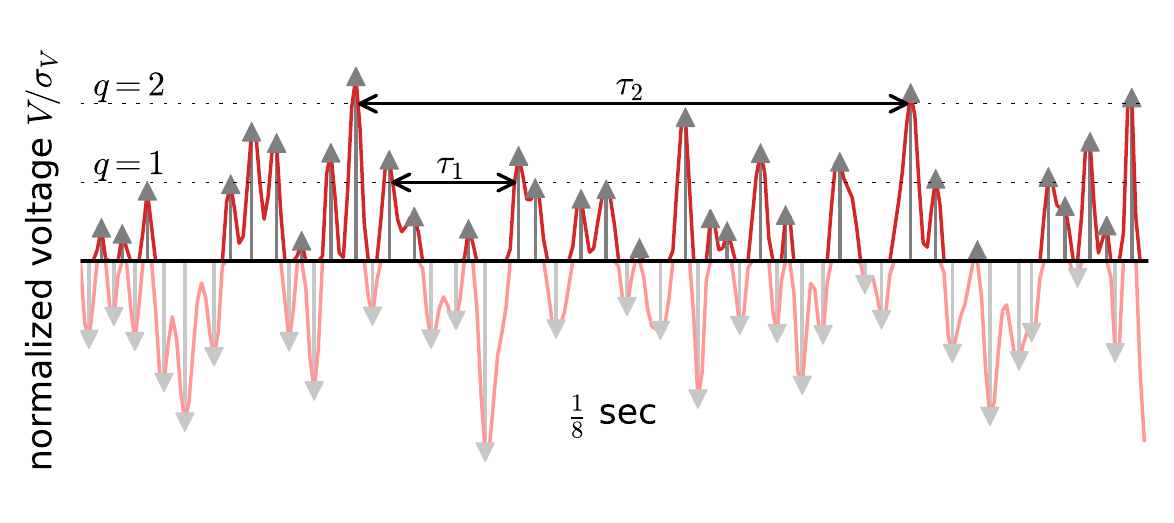} 
\caption{{\bf Schematic illustration of inter--pulse intervals}. Shown are the
normalized data (red) together with the related point process (gray) for \nicefrac{1}{8} 
second of total 10 seconds of recording. 
The black line illustrates the baseline signal, here at $q=0$.
Triangles directed upwards (downwards) mark the positive maxima (negative minima). Arrows depict exemplary
IPIs $\tau_1$ and $\tau_2$ for the respective threshold values 
$q=1$ (bottom) and $q=2$ (top).}
\label{fig1}
\end{figure}
The process consists of positive maxima (negative minima) created by the electrical activities of MU,
cf. Fig. \ref{fig1}. In the following we limit to the description of the 'upper' part of the signal
where $V \ge 0$.
Exactly the same procedure can be adopted for 'lower' negative fraction, thus
gives the improvement of the statistics.
The maxima above the baseline were used in the construction of the point process.
As each maximum is replaced by the point of the same height as the analogous 
maximum and all other values
of the signal are set to zero the resulting time train of discrete spikes constitute a
point process which can be described by the random intervals between the points. 
The archetype of the point process is well known Poisson process (or Poisson point field)
which has the property that each point is stochastically independent of all the other 
points in the process \cite{Daley2003}. As such, the distribution of the intervals between
the consecutive points is known to be exponential. Despite being a limiting case of Bernoulli trials, 
the process was shown to have numerous applications in distant branches of science, including
astronomy \cite{babu1996spatial}, biology \cite{othmer1988models}, image processing
\cite{bertero2009image}, finance markets \cite{yamasaki2005scaling}, or even for classical
\cite{Spiechowicz2013} and quantum \cite{luczka1991master} dynamical systems,
to name but a few.

As we are interested in statistical properties of the signal at all possible scales of 
volatilities, small, intermediate 
and large, the range of thresholds $q$ were set from 0 to the maximum possible
value of rescaled voltage, above which at least $10$ maxima were found. This will
allow for the description of the properties of the whole signal, regardless the scale.
The maximum value 
of the threshold would differ for each channel, depth and stage of treatment for every 
analyzed patient. Nevertheless, the threshold was found never to exceed $q=10$.

\subsection*{Data acquisition}
Signals were obtained with use of reusable anal probe developed at Laboratory of Engineering of Neuromuscular System and Motor Rehabilitation of Politecnico di Torino in collaboration with the company OT-Bioelettronica. The probe thickness is 14 mm and it has 3 rings of 16 silver bar electrodes each. Electrodes are placed parallel to the long axis of the probe and have dimensions of $9 \times 1$ mm. There is an 8 millimeter distance between each of the rings hence they enable recordings at 3 different levels of anal canal. Probe was positioned in a way that first ring would be placed starting from the anal verge, second would start at 18 mm and third one at 35 mm. 
The sampling frequency was 2048 Hz, which for the 10 seconds of the measurement gave 20480 data points. The probe was connected to the standard PC over 12 bit NI DAQ MIO16 E-10 transducer (National Instruments, USA). Low and high pass filters were used at 10 and 500 Hz respectively which gave a bandwidth equal to 3 dB. The monopolar signals were obtained by differential acquisition of the signals detected from each pair of the electrodes of the probe with respect to a reference strip connected to the patient wrist \cite{Cescon}. For the time of the acquisition patients were positioned in left lateral decubitus position. Measurement were conducted in a constant pattern: 1 minute relaxation, three 10 sec long recordings at rest for each depth, 1 minute relaxation and then three 10 sec long recordings at maximum voluntary contraction (MVC) for each depth with additional 1 minute breaks in between. The whole process was repeated twice, which gave 12 recordings for a patient for each time point. The recordings were conducted at four time points of treatment - before the surgical procedure ($D_1$) and 1 month ($D_2$), 6 months ($D_3$) and 1 year ($D_4$) after the surgery.

\subsection*{Patients}
The study included 16 subjects, 5 female, age range 46 to 71 (average 56 years) and 11 male, age range 40 to 85 (average 63,6 years), diagnosed with rectal cancer and qualified for surgery. Based on localization of rectal cancer patients underwent either Low Anterior Resection (LAR) - 9 patients, Anterior Resection (AR) - 6 patients or proctocolectomy (PC) - 1 patient. The detailed information about the surgery of the rectal cancer and the role of sEMG for the patient diagnosis can be found in \cite{delaini2005functional, lopez1999electromyography}.

\subsection*{Analysis}
The statistical properties of the inter--pulse intervals (IPIs) between consecutive maxima (see 
$\tau_1$ and $\tau_2$ in Fig \ref{fig1}) of the sEMG signals is the subject of this work.
Thresholds at which the intervals were calculated were determined in the logarithmic manner
from $0$ (effectively taken all of the maxima) 
up to the value where at least $10$ IPIs were evident. Keep
in mind that thresholds were also scaled with the standard deviations $\sigma_V$ complementary
to the signals. 
\begin{figure}[htbp]
\includegraphics[scale=1]{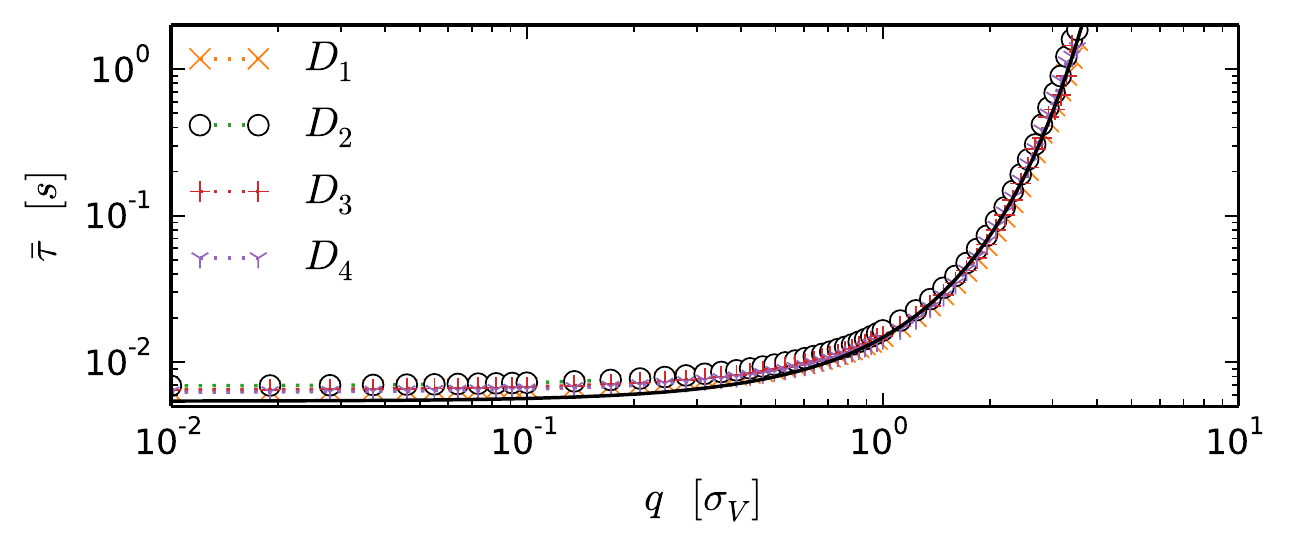} 
\caption{\textbf{The generic characteristics of the average IPI time $\bar{\tau}$ \textit{versus}
the threshold magnitude $q$ presented for all stages $D_1$ -- $D_4$ of treatment. }
Please note almost identical relation for all stages. Maximal threshold value for which at least $10$
IPIs were found does not reach $q=4$. The black line presenting fitted exponent is shown to 
lead the eye.}
\label{fig2}
\end{figure}
The typical dependence of the mean IPI time $\bar{\tau}$ on threshold $q$
is presented
in Fig \ref{fig2} for the selected patient and for all stages $D_1$ -- $D_4$ of treatment. As expected
it is the monotonic function of the threshold value with the evident exponential growth.
The presented behavior 
seems to be the typical situation for the data registered for all patients. It means, that
the knowledge on the average IPI will not help to classify the stage of treatment. 
Maximal threshold values for which at least $10$ IPIs were found do not reach $q=4$
for most examined cases. For small values of the threshold the mixed statistical 
properties of MUAPs, the background noise, failed initiations of MUAPs as well as 
refractory periods of 
MUAPs or satellite waves are all taken into consideration. For intermediate levels of $q$
one can expect only MUAPs and satellite waves to contribute to the effective, above the
threshold train of pulses. For large thresholds, $q > 2$, the average waiting times
between consecutive maxima grows by the order of magnitude therefore are likely to 
portray only the properties of MUAPs. 
\begin{figure}[htbp]
\includegraphics[width=1.\linewidth]{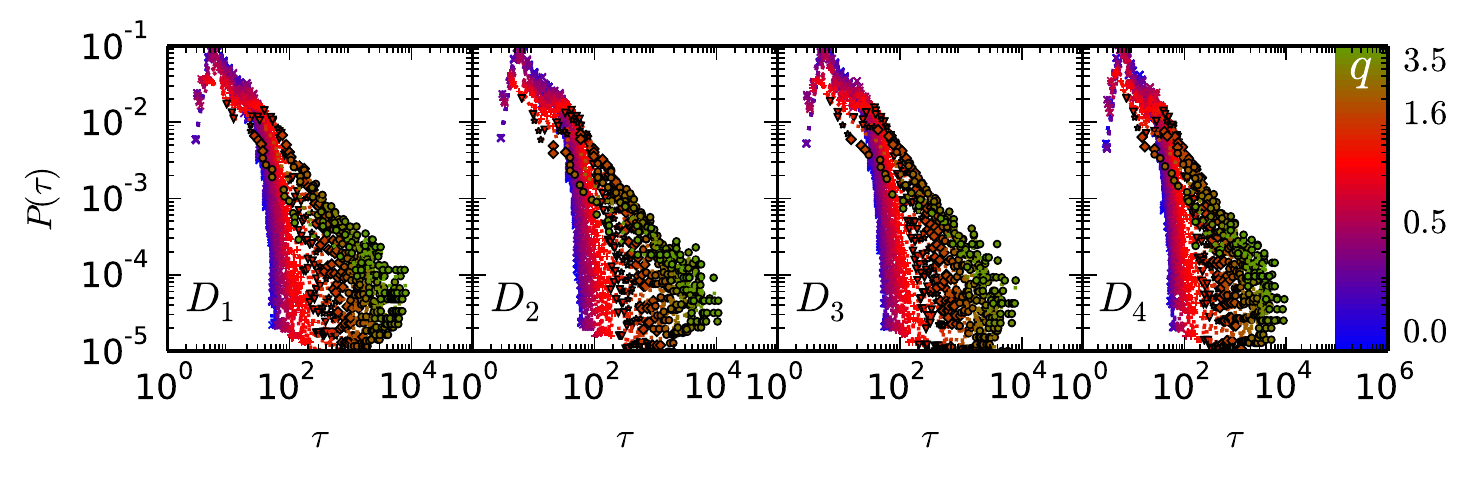}
\caption{{\bf Probability distribution function $P(\tau)$ presented for several 
threshold values $q$.} Computed PDFs cannot be represented by one specific function.
The PDF for the threshold level are indicated by the colour 
-- see colour bar on the right.}
\label{fig3}
\end{figure}
Next, we study the behavior of the probability distribution function (PDF) and it's 
dependence on the threshold value $q$. For different thresholds PDFs are different cf. 
Fig \ref{fig3} and cannot be represent by the same function. In particular they cannot 
be described by the Poisson distribution $\propto \exp(-\tau)$ as for the uncorrelated 
data. This in turn allows to draw a conclusion of existing correlations hidden in the 
acquired data. To 
shed some more light on the actual influence of the threshold value the scaled PDFs 
$P(\tau) \bar{\tau}$ as a function of the scaled inter--pulse interval $\tau / \bar{\tau}$
is presented in Fig \ref{fig4}. Again the different colors, blue -- purple -- red -- 
orange -- green, reflect the growing threshold dependence, from $q=0$ up to $q \simeq 4$.
Different symbols group the IPIs for different values of the average
IPI $\bar \tau$, with $\times$ being equivalent to average time shorter than $0.01$ sec 
and $\circ$ longer than $0.1$ sec. As one can notice the scaled PDF $P(\tau) \bar{\tau}$ 
dependence on scaled IPI is found to be different both for different 
average times (marked by symbols) and thresholds (marked by colors). Keep
in mind an exponential correspondence between $\bar \tau$ and $q$, 
cf. Fig \ref{fig2}. 
\begin{figure}[htbp]
\includegraphics[scale=1]{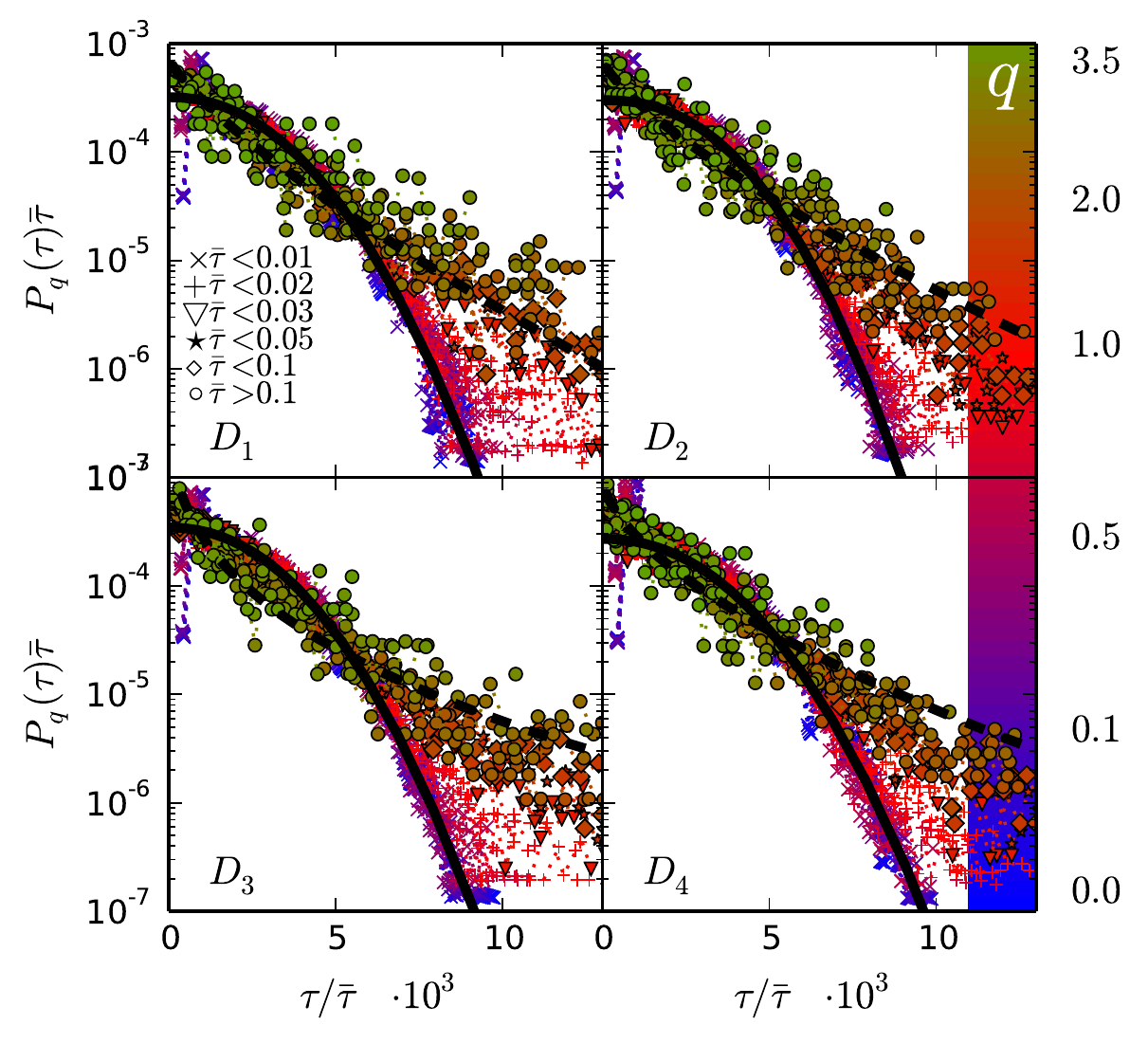}
\caption{{\bf Probability distribution function scaled with the average IPI time $\bar{\tau}P(\tau)$
\textit{versus} scaled IPI time $\tau / \bar{\tau}$.} Different symbols were used in order to provide
the visual guidance for different average IPIs -- see legend in top right panel ($D_1$).
For all the stages $D_1-D_4$ one can reveal the stretched--exponential function (\ref{sexp})
with the threshold $q$ dependent parameters.
Again the PDFs for the different threshold $q$ level are indicated by the colour -- 
see colour bar on the right.}
\label{fig4}
\end{figure}

In spite of the visual differences of the scaled probability distribution function
\begin{equation}
P_q(t) = \frac{1}{\bar \tau} f_q\left(\frac{t}{\bar \tau} \right)
\label{pdf}
\end{equation}
for different average IPIs or thresholds, equivalently, the analysis is consistent
with the possibility that the data can be characterized by the stretched exponential 
function 
\begin{equation}
f_q(x) \propto \exp(-b_q x^{\beta_q})
\label{sexp}
\end{equation}
where all the parameters $\{b_q, \beta_q\}$ are threshold dependent.

\subsection*{Correlation between records}
\begin{figure}[htbp]
\includegraphics[width=1.\linewidth]{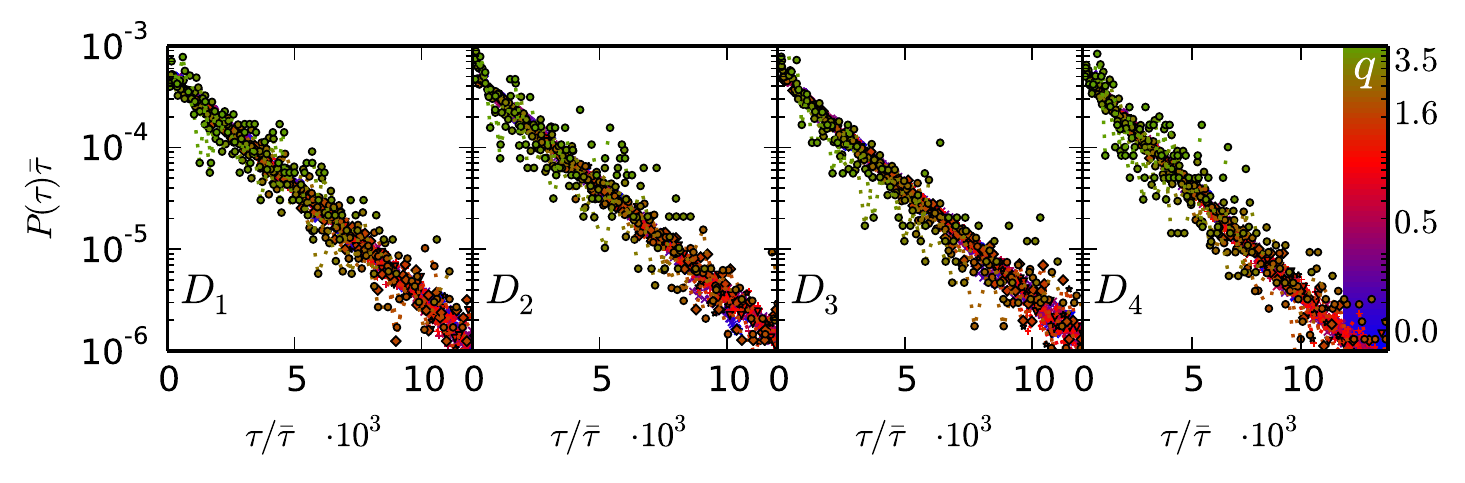}
\caption{{\bf Probability distribution function for the shuffled IPIs.}
For all the stages $D_1$--$D_4$ one can reveal the exponential Poisson-like
distribution $\log[P_q \bar \tau] = -\tau/\bar{\tau}$, seen as a linear function 
on the semi-logarithmic plot.}
\label{fig5}
\end{figure}

\begin{figure}[htbp]
\includegraphics[scale=1]{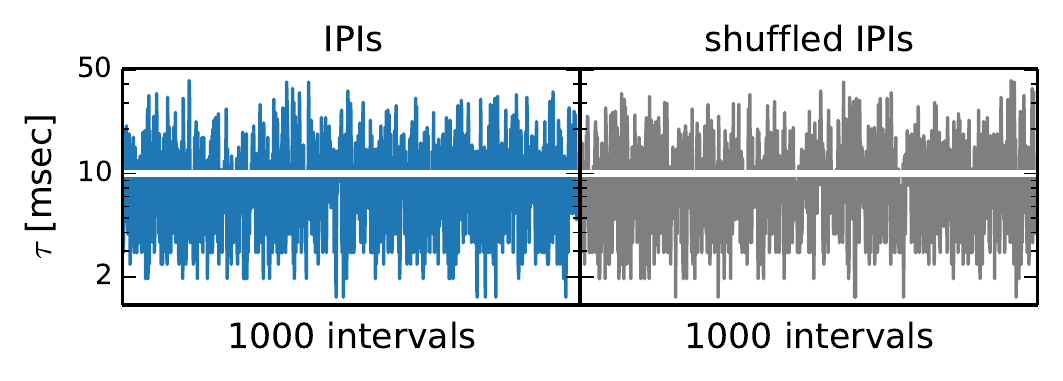}
\caption{{\bf Typical train of the analyzed inter--pulse intervals.}
On the l.h.s. the original train of the $D_1$ stage of one of the patients
is presented for the average inter--pulse interval $\tau = 0.01$ msec. On the
r.h.s. shuffled data is shown.}
\label{fig6}
\end{figure}

\begin{figure}[htbp]
\includegraphics[width=0.99\linewidth]{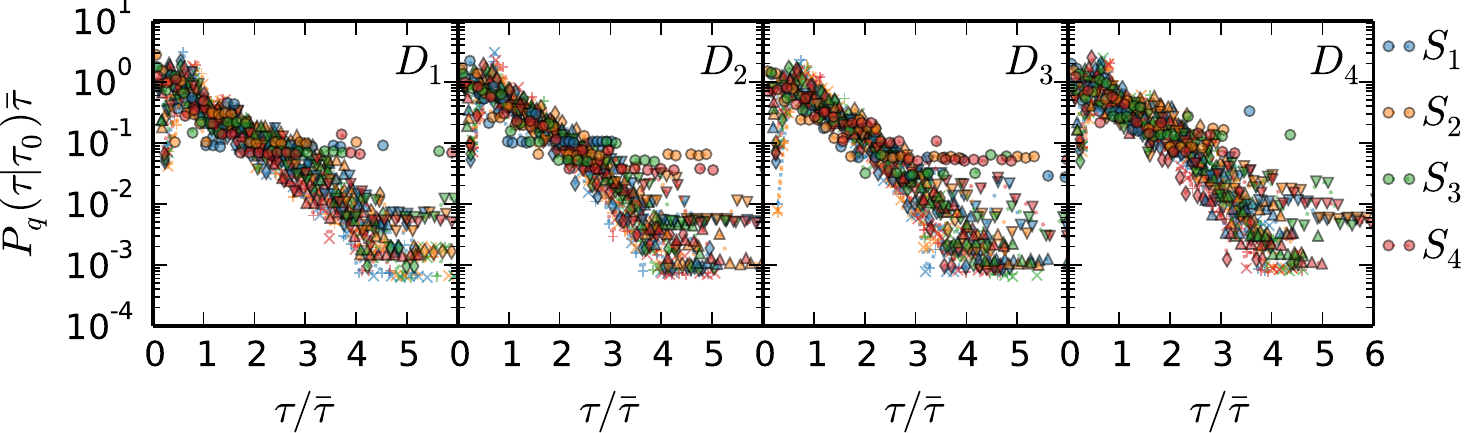}
\caption{{\bf Conditional probability distribution function for IPIs.}
For all the stages $D_1$--$D_4$ one can reveal the exponential Poisson-like
distribution $\log[P_q(\tau | \tau_0) \bar{\tau}] = -\tau/\bar{\tau}$, seen as a linear function 
on the semi-logarithmic plot.}
\label{fig7}
\end{figure}

For uncorrelated records one expects the data to follow the exponential Poisson
distribution. Here we claim, that the analyzed IPIs do possess the correlations.
To test this expectations and prove the long-term memory we shuffle the inter--pulse
intervals $\tau$ and investigate the resultant PDFs. The shuffled data indeed
follow the Poisson distribution, see Fig \ref{fig5}. This form of the PDF for the shuffled
IPIs suggest that the stretched exponential form of the distribution function for the
original IPI emerges from the correlations.
The closer look at the inter--pulse intervals trains of original and shuffled 
data are rather similar, cf. Fig \ref{fig6}. There are also no visible clustering, or any other
indications of memory -- the events look simply random.
If the distribution (\ref{pdf}) suppose to fully characterize the underlying data, the subsequent IPIs 
should not be dependent on each other, and only be chosen randomly from the distribution 
(\ref{pdf}). In turn, the obtained PDF would fully characterize the data. 
To test the possible effects of memory among the intervals we study the conditional probability
distribution function $P_q(\tau | \tau_0)$ of the intervals $\tau$ which directly follow the interval
$\tau_0$. In order to do so first we sort the IPIs in the ascending order. Next we split just sorted
intervals into four equinumerous subsets $\{S_1, S_2, S_3, S_4\}$. The shortest intervals will fall 
into the first set $S_1$ while the longest intervals will be stored in the last one $S_4$. The results
shown in Fig \ref{fig7} demonstrate the same exponential behavior of the conditional PDF for any
threshold $q$ regardless the set the $\tau_0$ was taken from. In Fig \ref{fig7} one can find
plots of the conditional PDF $P_q(\tau | \tau_0)$ scaled with the average interval $\bar \tau$
as a function of the scaled IPI $\tau / \bar{\tau}$.  
Four different colors blue, orange, green
and red were used to indicate the correspondent subset $\{S_1, S_2, S_3, S_4\}$ respectively. On top
of that one can track the behavior of conditional PDF for different threshold values $q$.
There is a striking correspondence for all subsets and thresholds. It means that there is no memory
effects between the consecutive intervals, also for the case of high thresholds. This in turn indicates
that the anterior MUAP has no influence on the posterior one.

% Results and Discussion can be combined.
\section*{Results and Discussion}
Lack of the memory effects in the IPI trains as well as Poisson like exponential shape of the 
PDFs for shuffled data let us state the hypothesis that the threshold dependent PDFs for the IPIs 
(\ref{pdf}) fully describe the behavior of the inter--pulse intervals of surface electromyography 
signals acquired from patients with colorectal cancer. Now we take a look at the actual shape
of the stretched exponential $f_q \propto \exp(-b_q x ^{\beta_q})$ (\ref{sexp}) .
\begin{figure}[htbp]
\includegraphics[width=0.5\linewidth]{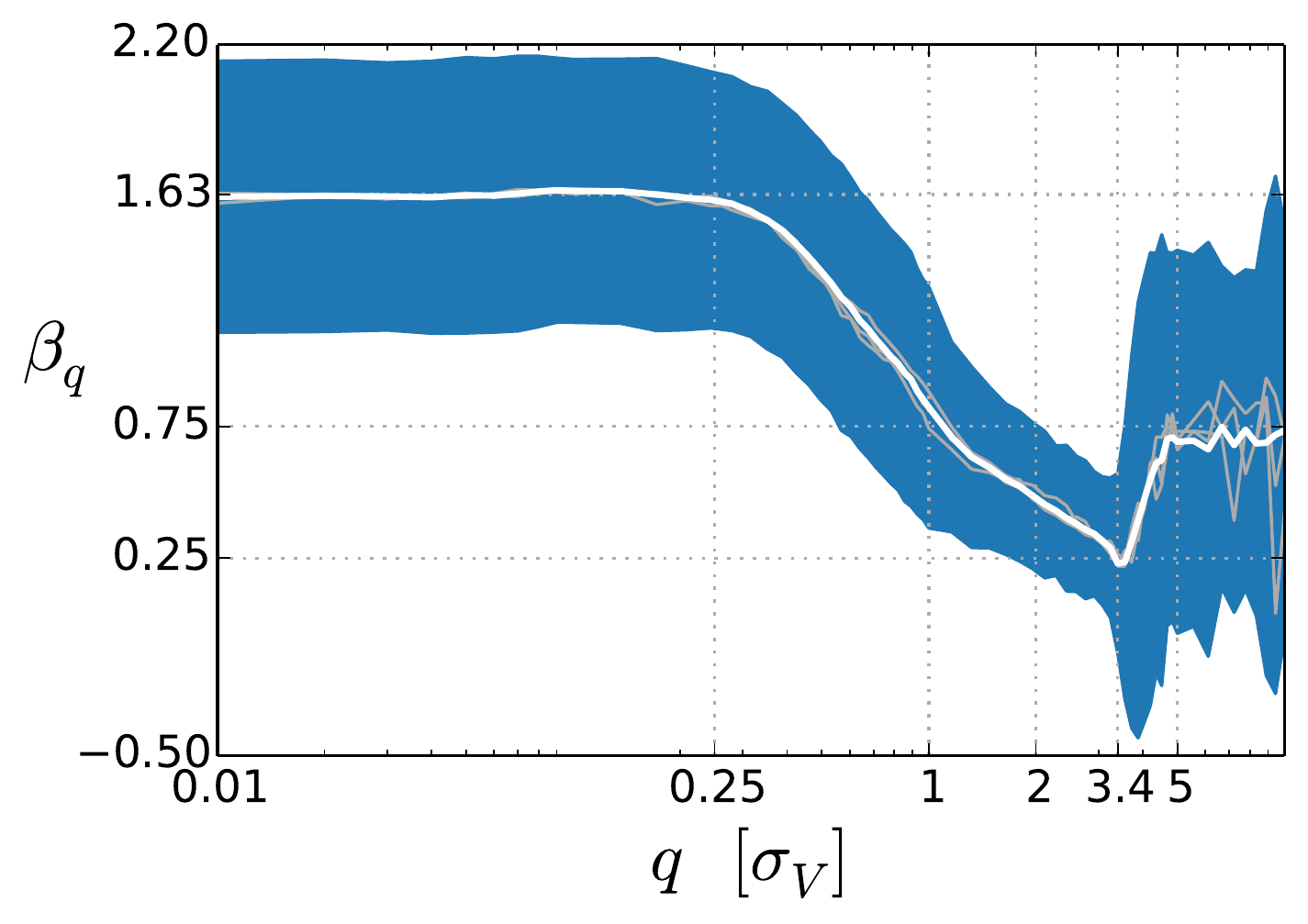} 
\caption{{\bf Average value of the shape parameter $\beta_q$ \textit{versus} the threshold level $q$.} 
shown at all four stages of treatment. The white line represent the global average calculated
for all possible cases $D_1$ -- $D_4$ and at both states MVC and at rest. Barely visible gray lines
which represent separate states $D_1$ -- $D_4$ shadow the global average. The dark blue band represents
the standard deviation $\sigma_{\beta_q}$.}
\label{fig8}
\end{figure}
The threshold dependence of the shape parameter $\beta_q$ is presented in the Fig \ref{fig8}. 
For all of the analyzed data the $q$ dependence can be characterised as a non--monotonic
function. As expected for relatively low values of threshold $q < 0.25 \sigma_V$ i.e. for the
situation of all the possible pulses analyzed the average
value of the shape parameter is constant and close to $\beta_q \simeq 1.62$. 
Next it decreases with increasing $q$ reaching a value of around $\beta_q \simeq 0.25$
for the threshold $q \simeq 3.4 \sigma_V$. From this value onwards it increases and saturates 
close to $\beta_q \simeq 0.75$ for thresholds higher than $5 \sigma_V$. The last part
corresponds to the highest pulses which can be treated as MUAPS.
This behavior is not dependent on the stage of treatment $D_1$--$D_2$. In the Fig \ref{fig8}
one can find gray lines which correspond to the consecutive stages and shadow the
average value. Up to the threshold
values close to $5 \sigma_V$ these are rather indistinguishable from each other and from
the average value of the shape parameter. Above this value the number of events drops 
drastically, thus the resulting statistics is weaker and the discrepancies
between stages and the average values are found. Nevertheless the tendency is similar and all the
curves fluctuate around the same average value.

From the above analysis it follows that the statistics of the inter--pulse intervals of the 
particular stages of colorectal cancer treatment are similar. The scaled probability distribution functions
$\bar{\tau}P_q(\tau)$ of the rescaled IPIs $\tau / \bar{\tau}$ exhibit stretched--exponential behavior
with the threshold (average IPI) dependent shape $\beta_q$ and scale $b_q$ parameters. The actual
PDF seems to be the sufficient characteristics for the description of the process. Existing long time correlations
purely comes from the distribution - the shuffled data show clean exponential nature. Clustering events
were also not detected for all of the stages. 

Based on our data we can conclude that analyzed parameters are not influenced by multimodal
treatment of rectal cancer hence the probability that aforementioned treatment affects inter--pulse
intervals between the extrema of an electrical activity of External Anal Sphincter is minimal.

\section*{Acknowledgments}
This work was partially supported by
the JUMC research grant: K/ZDS/006369.

\end{document}